# Stability Frontiers and Mixed-dimensional physics in the Kagome Intermetallics $Ln_3ScBi_5$ ($Ln$ = La-Nd, Sm)

Zhongchen Xu [1, 2, 3], Wenbo Ma [2, 3], Shijun Guo [2, 3], Ziyi Zhang [2, 3], Quansheng Wu [3], Xianmin Zhang [1, #], Xiuliang Yuan [2, 3, #], Youguo Shi [2, 3, 4, #]

[1] *Key Laboratory for Anisotropy and Texture of Materials (Ministry of Education), School of Material Science and Engineering, Northeastern University, Shenyang 110819, China*
[2] *Center of Materials Science and Optoelectronics Engineering, University of Chinese Academy of Sciences, Beijing 100190, China*
[3] *Beijing National Laboratory for Condensed Matter Physics and Institute of Physics, Chinese Academy of Sciences, Beijing 100190, China*
[4] *Songshan Lake Materials Laboratory, Dongguan, Guangdong 523808, China*

To whom correspondence should be addressed.
[*] zhangxm@atm.neu.edu.cn
[*] yuanxiuliang8@iphy.ac.cn
[*] ygshi@iphy.ac.cn

## Abstract

Low-dimensional physics provides profound insights into strongly correlated interactions, leading to enhanced quantum effects and the emergence of exotic quantum states. The $Ln_3ScBi_5$ family stands out as a chemically versatile kagome platform with mixed low-dimensional structural framework and tunable physical properties. Our research initiates with a comprehensive evaluation of the currently known $Ln_3ScBi_5$ ($Ln$ = La-Nd, Sm) materials, providing a robust methodology for assessing their stability frontiers within this system. Focusing on $Pr_3ScBi_5$, we investigate the influence of the zigzag chains of quasi-one-dimensional (Q1D) motifs and the distorted kagome layers of quasi-two-dimensional (Q2D) networks in the mixed-dimensional structure on the intricate magnetic ground states and unique spin fluctuations. Our study reveals that the noncollinear antiferromagnetic (AFM) moments of $Pr^{3+}$ ions are confined within the Q2D kagome planes, displaying minimal in-plane anisotropy. In contrast, a strong AFM coupling is observed within the Q1D zigzag chains, significantly constraining spin motion. Notably, the magnetic frustration is partially the consequence of coupling to conduction electrons via the Ruderman-Kittel-Kasuya Yosida (RKKY) interaction, highlighting a promising framework for future investigations into mixed-dimensional frustration in $Ln_3ScBi_5$ systems.

INTRODUCTION

In low-dimensional physical systems, the restricted paths for conducting electrons enhance interactions between particles, leading to strongly correlated interactions and enhanced quantum effects that are pivotal in the development of semiconductor devices[1], topological insulators and quantum computing[2, 3], as well as superconductivity research[4]. The understanding of the strong correlation mechanism in Q1D materials, which consist of one-dimensional (1D) atomic chains or columnar moieties, is rooted in the Luttinger Liquid theory[5], describing electron behavior in 1D conductors. Within these materials, Q1D spin chains coexist with inter-chain coupling, offering a versatile platform for investigating the magnetism of various models, including Ising[6], Heisenberg[7], and XY-type models[8]. Moreover, highly frustrated Q2D kagome lattice magnet are characterized with geometric frustration and unique electronic structures, which can give rise to various intriguing quantum states. Vanadium-based kagome materials $A$V$_3$Sb$_5$ ($A$ = K, Rb, Cs) a range of exotic phenomena, including superconductivity, anomalous Hall effect and electronic nematicity [9-11]. The distorted kagome intermetallic compound CePdAl manifests a paramagnetic quantum-critical phase due to the competitive interplay of geometric frustrations, RKKY interactions, and Kondo effects[12]. The exploration of complex magnetic ground states and quantum transport arising from the interplay of Q1D and Q2D hybrid low-dimensional characteristics in a system is an intriguing endeavor.

The $Ln_3$ScBi$_5$ family ($Ln$ = La - Nd, Sm) has attracted our research interest, as it simultaneously hosts Q1D zigzag spin chains and Q2D distorted kagome motifs and making it an excellent candidate for studying mixed-dimensional frameworks. The hypervalent bismuthide La$_3$ScBi$_5$ is distinguished by its unique 1D bismuth chains, which has given rise to unusual phenomena including quasi-linear magnetoresistance, quantum oscillations, and paramagnetic singularity[13]. The frustrated Q2D Kondo-lattice compound Ce$_3$ScBi$_5$ exhibits interactions with crystal electric field (CEF) splitting, Kondo scattering and RKKY interaction at closely related energy scales, leading to the emergence of fractional magnetization plateaus and singular anisotropic transport[14]. For the $Ln_3$ScBi$_5$ system, it is of significant physical importance to explore the synthesis of heavier rare earth elements to investigate the stability frontiers and to analyze the common features induced by the mixed-dimensional structure across the system.

In this study we successfully optimized the flux-growth synthesis conditions and systematically synthesized a series of high-quality single crystals exhibiting *anti-Hf$_5$Sn$_3$Cu*-type hexagonal structures from the $Ln_3$ScBi$_5$ family. Through a comprehensive analysis, the structural diversity and stability boundaries within this family were visualized and evaluated. Special attention is given to Pr$_3$ScBi$_5$, where we investigate the influence of the zigzag chain of Q1D and the kagome layer of Q2D in the mixed-dimensional structure on its magnetic ground state. The ordered magnetic moments of the non-collinear Pr$^{3+}$ ions are confined to the Q2D kagome plane, exhibiting minimal in-plane anisotropy, while strong AFM coupling is observed in the Q1D zigzag chains, where spin motion is significantly constrained. These unique features can also be extended to the $Ln_3ScBi_5$ family. It is worth noting that the magnetic entropy at the AFM transition is significantly reduced, and the transition occurs at a lower temperature below the wide peak of magnetic susceptibility, indicating pronounced spin fluctuations in $Pr_3ScBi_5$. The $Ln_3ScBi_5$ system serves as a paradigm for further investigating the interplay between spin frustration and RKKY interaction across mixed dimensions.

METHODS

**Optimized flux single-crystal growth.** Millimeter-scale single crystals of $Ln_3ScBi_5$ were synthesized using a self-flux method. Rare earth ($Ln$) lumps were physically polished to remove the oxide layer, then weighed together with scandium (Sc) pieces and bismuth (Bi) balls at a certain molar ratio. The precise ratio of $Ln$: Sc: Bi is flexible; Reproducible and optimal ratios were experimentally determined: Pr: Sc: Bi = 1:4:10, Nd: Sc: Bi = 1:4:10, Sm: Sc: Bi = 1:3:5. The reagents were placed in a 5 ml $Al_2O_3$ crucible containing a connecting crucible, sealed in a fused silica tube under approximately 0.6 atm of argon gas. The samples were heated to 1000 – 1050 °C over 10 hours, held at this temperature for 18 hours, then slowly cooled at a rate of 1 – 1.5°C/h to 700 °C, followed by centrifugation at 700 °C.

$Ln_3ScBi_5$ single crystals exhibit a typical hexagonal morphology. Residual Bi flux on the crystal surfaces was removed using a scalpel, revealing a lustrous silver appearance. Crystal size is constrained by the volume of the growth container. Members of the $Ln_3ScBi_5$ family yield samples with average centimeter-scale dimensions, with the exception of $Sm_3ScBi_5$, which forms slender filaments or strips, with an average size of only $0.5 \times 0.5 \times 5$ mm$^3$. These single crystals exhibit air sensitivity, necessitating both their subsequent storage and preparation to be conducted within an argon-filled glove box.

**Characterizations.** The diffraction peaks corresponding to the ($h$00) surfaces of single crystals were conducted using a Bruker D2 phaser detector with Cu K$\alpha$ radiation. Single crystals of $Ln_3ScBi_5$ were mounted on kapton loops with Paratone oil for single-crystal X-ray diffraction (XRD) data collection. Diffraction data were collected at 274(2) K on a Bruker D8 VENTURE PHOTON II diffractometer using Mo K$\alpha$ radiation ($\lambda$ = 0.71073 Å). Structure solutions were obtained using intrinsic phasing methods with the APEX4 software, and the crystal structure was refined using a full-matrix least-squares method on $F^2$ with the SHELXL-2018/3 program. The chemical composition of the as-grown crystals was determined using energy-dispersive X-ray spectroscopy (EDS) in a Hitachi S-4800 scanning electron microscope at an accelerating voltage of 15 kV.

**Physical properties measurement.** Temperature-dependent susceptibility was conducted from 2 to 300 K using a Magnetic Properties Measurement System (MPMS, Quantum Design) with varied applied fields in the zero-field-cooling (ZFC) and field-cooling (FC) modes. Isothermal magnetization was assessed on the MPMS (7 Tesla) and a Physical Properties Measurement System (PPMS, Quantum Design, 16 Tesla) utilizing a vibrating sample magnetometer (VSM) option. Electrical transport measurements were performed on the PPMS (up to 9 Tesla) using the standard four-probe technique built by platinum wire and silver epoxy with current applied along c-axis. The resistance signals were obtained by symmetrizing the data collected in reversed magnetic fields. Heat capacity was measured in the PPMS using its specific heat option, employing a thermal relaxation method. The addenda data were measured in advance, and the temperature-rise parameter was set to a minimal value of 1% to achieve the best temperature resolution.

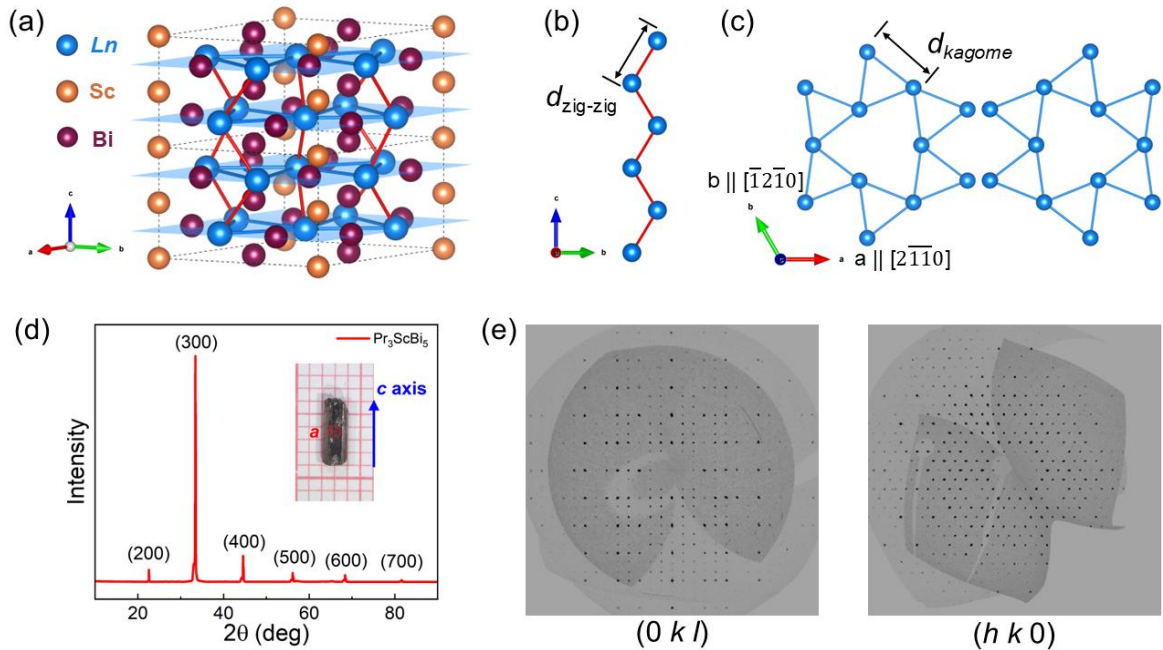

Figure 1. Schematic crystalline structure of the $Ln_3ScBi_5$ family, utilizing $Pr_3ScBi_5$ single crystal as an example for characterization. (a) Overall crystal structure of $Ln_3ScBi_5$ showcases quasi-one-dimensional (Q1D) zig-zag chains along the $c$-axis direction (b), exhibiting quasi-two-dimensional (Q2D) distorted kagome lattice with the $ab$ plane, and a stacking arrangement along the crystallographic $2_1$ screw axis (c). (d) XRD $\theta$-$2\theta$ scan of a flat surface of $Pr_3ScBi_5$ single crystals. (e) Reconstructed single-crystal XRD patterns of $(0kl)$ and $(hk0)$ reflection planes, respectively.

## RESULTS AND DISCUSSION

***$Ln_3ScBi_5$ Structural Trends.*** $Ln_3ScBi_5$ crystallizes in an *anti-$Hf_5Sn_3Cu$*-type hexagonal structure, which can alternatively be interpreted as an interstitial derivative of the *$Mn_5Si_3$-type* structure. In this framework, the rare-earth atom occupies the single Wyckoff site 6$g$, Sc resides the site 2$b$, and Bi is distributed across two distinct sites: Bi1 at 6$g$ and Bi2 at 4$d$. This specific arrangement leads to an anisotropic environment where Sc is coordinated by octahedra of Bi1, which share faces along the hexagonal $c$ [0001] direction (Fig. 1a). The structure of $Ln_3ScBi_5$ exhibits a unique mixed-dimensional characteristic, featuring Q1D zig-zag chains of rare-earth atoms aligned along the $c$-axis (Fig. 1b) and Q2D distorted kagome layers, symmetrically stacked through the 6$_3$ screw axes of Sc atoms within the $ab$-plane (Fig. 1c). The incorporation of such mixed-dimensional elements, together with the coexistence of rare-earth metal magnetism, can foster a rich spectrum of exotic quantum phenomena.

Crystals of $Ln_3ScBi_5$ grown from a bismuth self-flux exhibit a bright, silver luster and are mechanically soft, allowing for easy cleavage. As an illustrative example, further characterization and analysis were performed on a single crystal of $Pr_3ScBi_5$. Figure 1d displays the XRD pattern of single crystal plane, where ($h$00) peaks are prominently observed, and the rod-shaped sample extends along the $c$-axis. Figures 1e and 1f present reconstructed single-crystal XRD patterns of (0$kl$) and ($hk$0) reflection planes, respectively, measured at 274 K. Comprehensive details regarding CIF files, refinement parameters, and additional characterization are provided in Table S1-S3 and Fig. S1 of the Supporting Information.

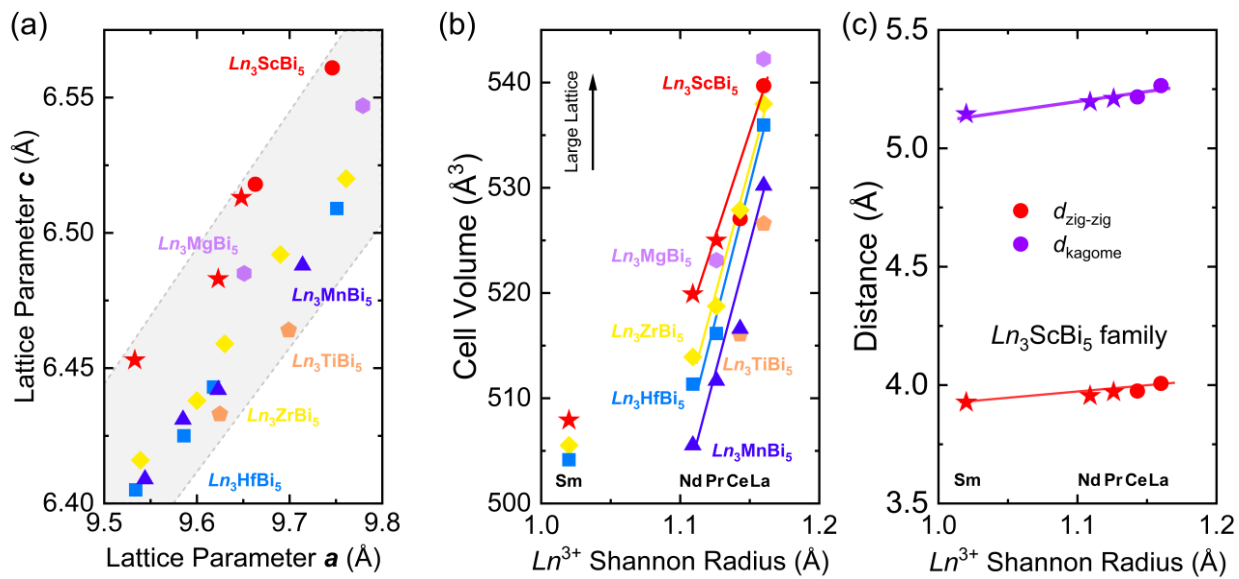

Figure 2. Stability frontiers of $Ln_3ScBi_5$ family. (a) Plot of the $a$ and $c$ lattice parameters for all currently known $Ln_3MBi_5$ ($Ln$ = La -Nd, Sm, $M$ = Sc, Mg, Ti, Zr, Hf) compounds[13-24]. The five-pointed star symbolizes the recently synthesized compound in this study, signifying an expansion of the system. (b) Schematic of phase stability across the lanthanide row. The coordination of $Ln$ is approximately VIII-coordinate in the $Ln_3MBi_5$ structure. (c) Nearest-neighbor distance between Nd atoms in zig-zag chains and kagome layers in $Ln_3ScBi_5$ family, extracted from SCXRD results, plotted against the VIII-coordinate Shannon ionic radius.

The intricate connection between the size of the filler atoms and host lattice results in their robust structure and tunable diversity within the $Ln_3ScBi_5$ and related systems. To illustrate this, the lattice parameters $a$ and $c$ for known $Ln_3MBi_5$ compounds have been plotted, as shown in Fig. 2a[13-24]. The overall trend is visually striking, the $Ln_3ScBi_5$ family exhibits an extended lattice parameter $c$, which can be attributed to the larger combined metal radii of the Sc and Bi atoms within their octahedral configurations. Furthermore, the lattice parameter $a$ of $Ln_3ScBi_5$ is slightly smaller than that of $Ln_3MgBi_5$, resulting in a similar unit cell volume between the two systems (Fig. 2b). In contrast to the usual lanthanide trend, when rare earth atoms shrink, the $Ln_3MBi_5$ series deviates from the typical linear trend observed in rare earth substitution series [25]. As smaller rare earth elements are incorporated, the rate of volumetric contraction within the lattice progressively diminishes, which is attributed to the resistance of the M-Bi bond within the octahedron to further compression.

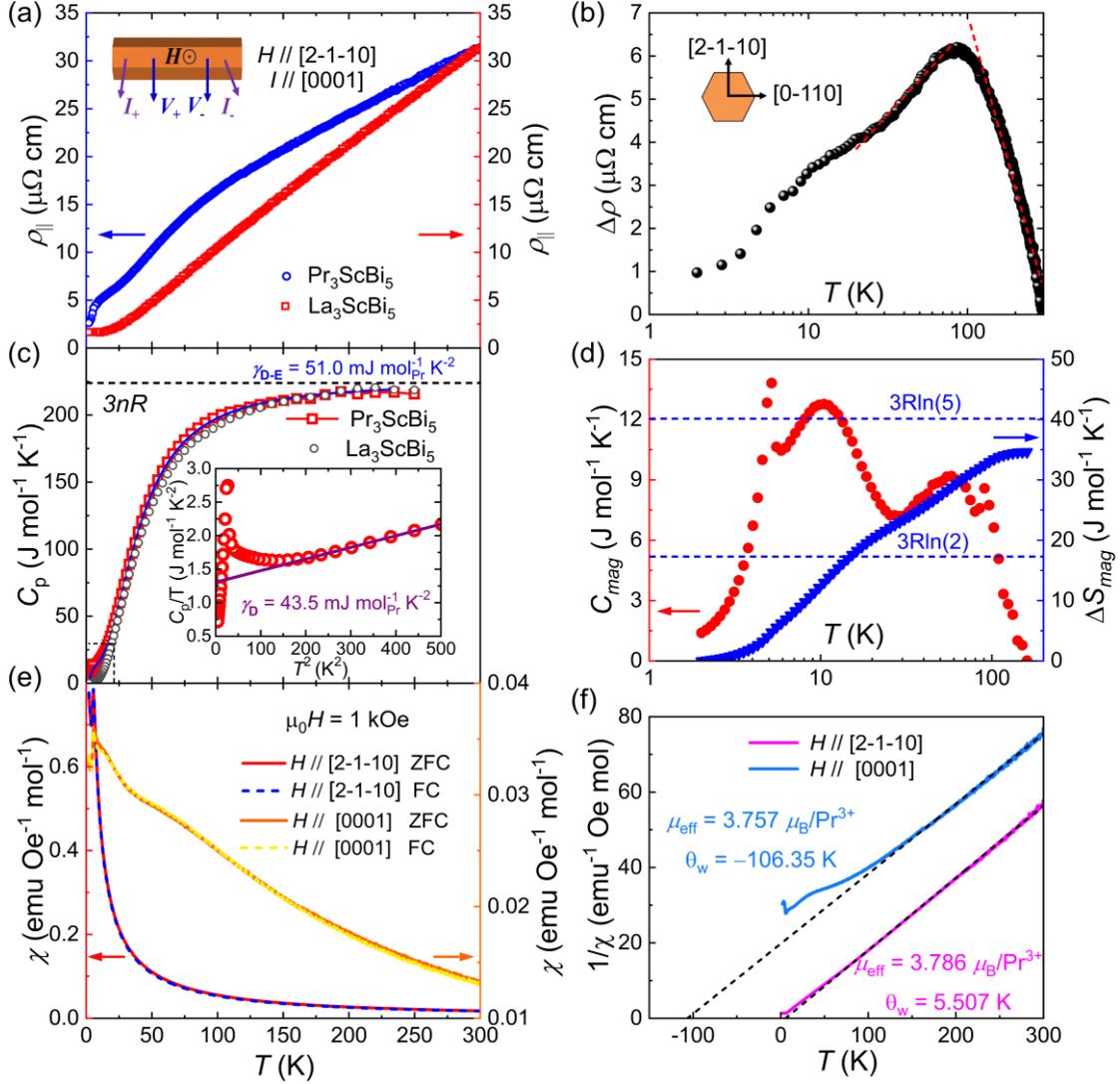

Figure 3. Summary chart of temperature-dependent physical properties of $Pr_3ScBi_5$. (a) Temperature-dependent resistivity of $Pr_3ScBi_5$ and its nonmagnetic analog $La_3ScBi_5$ in zero field, measured for the current flowing along $c$-axis. (b) Resistivity induced by the magnetism of $Pr^{3+}$ ions. The dashed straight line indicates the logarithmic slop of the resistivity. (c) Temperature-dependent specific heat of $Pr_3ScBi_5$ and $La_3ScBi_5$. The blue curve represents the Debye-Einstein fitting. The inset depicts the low-temperature $C_p/T$ vs $T^2$ behavior, with the purple line representing the Debye model linear fit. (d) Specific heat and entropy associated with the magnetic component. (e) Temperature variation of magnetic susceptibility measured in a magnetic field of $\mu_0H = 1$ kOe applied parallel $[2\bar{1}\bar{1}0]$ ($[100]$ $a$-axis) and $[0001]$ ($[001]$ c-axis) directions. Zero-field cooling (ZFC) and field cooling (FC) data are shown by solid line and dashed lines, respectively. (f) Curie-Weiss fit to the magnetic susceptibility data.

Through systematic and comprehensive experimental investigations, the stability boundary of $Ln_3ScBi_5$ can be elucidated from multiple perspectives with enhanced depth. When the Shannon radius of rare earth is reduced to that of $Sm^{3+}$ ions, the synthesis of high-quality $Sm_3ScBi_5$ single crystals becomes highly challenging: the yield of single crystals is extremely low, and they predominantly form long, slender filaments. This is attributed to the significant radius difference between charge balancing sublattice, which increases the structural instability and hampers single crystal growth[26], as corroborated by the volume deviation of $Sm_3ScBi_5$ from the linear trend in Fig. 2b. When employing smaller rare earth species such as $Gd^{3+}$, $Er^{3+}$, and $Lu^{3+}$, the preferential precipitation of the $Ln_3ScBi_5$ phase is inhibited, leading to the predominant formation of a simple rock-salt structure featuring Sc-doped $Ln_{1-x}Sc_xBi$.

Figure 2c makes a detailed comparison of the nearest neighbor distances within the zig-zig chains of Q1D motifs and within the distorted kagome layers of Q2D networks in the $Ln_3ScBi_5$ system. Despite the significant variation in the Shannon radius of rare earth, the responses of $d_{zig\text{-}zag}$ and $d_{kagame}$ in $Ln_3ScBi_5$ are comparatively modest, with both $d_{zig\text{-}zag}$ being substantially shorter than $d_{kagame}$, which indicates that the nearest neighbor distance of rare earth elements falls within the Q1D zig-zag chains. The RKKY interaction mediates long-range magnetic exchange in the intermetallic compound $Ln_3ScBi_5$, potentially stabilizing a diverse range of quantum ground states within its geometrically frustrated Q2D lattices[27]. By employing $Pr_3ScBi_5$ single crystals as a design paradigm, we explore the exotic spin fluctuations and diverse magnetic ground states induced by the complex structural mixed dimensionality in the $Ln_3ScBi_5$ system.

**Frustrated kagome-lattice Antiferromagnets.** The summary chart of the physical properties of $Pr_3ScBi_5$ varying with temperature is shown in Fig. 3, highlighting a strong correlation between electrical transport and magnetic order. The electrical resistivity $\rho_\parallel(T)$ of $Pr_3ScBi_5$ for $H$ // $[2\bar{1}\bar{1}0]$ decreases monotonically with decreasing temperature, showing metallic behavior (Fig.3a). The large residual resistivity ratio (RRR = $\rho_{\parallel(300\ K)}/\rho_{\parallel(2\ K)}$ ≈11.9) suggests that the as-grown crystals are of excellent quality and that the results are reproducible. The magnetic component of electrical resistivity $\rho_{mag}(T)$, estimated by subtracting phonon contribution using the $\rho_\parallel(T)$ data of $La_3ScBi_5$ (Fig. 3b), exhibit a $-\log T$ dependence in the temperature ranges 110 K ≤ T ≤ 300K and 20 K ≤ T ≤ 90K, and shows the emergence of the broad hump around 100 K. This behavior can be attributed to Kondo-type interaction in the presence of strong CEF splitting, with the Kondo temperature being much lower than the overall crystal field splitting [28, 29]. Caution is warranted when attributing resistivity changes to magnetic effects via curve subtraction, as impurity scattering variations between samples may obscure the interpretation. Notably, the resistivity undergoes a significant decrease below $T_N$ = 5.1 K, indicating the onset of magnetic ordering in $Pr_3ScBi_5$, which results in a reduction of magnetic scattering. The bulk nature of the magnetic transition observed in transport measurements is confirmed by the λ-shaped peak with appreciable magnitude near $T_N$ in the $C_P(T)$ curve. The zero-field $C_P$ data in the temperature range 1.8-200 K were analyzed using a combination of the Debye and Einstein models[30] according to:

$$C_p = \gamma^{D-E}T + 9nR\alpha\left(\frac{T}{T_D}\right)^3 \int_0^{\frac{T_D}{T}} \frac{x^4 e^x dx}{(e^x - 1)^2} + 3nR(1-\alpha)\left(\frac{T_E}{T}\right)^2 \frac{e^{\frac{T_E}{T}}}{\left(e^{\frac{T_E}{T}} - 1\right)^2}$$

The optimal fitting the yields parameters $\gamma^{D-E}$=51.0 mJ $mol_{Pr}^{-1}K^{-2}$, $T_D$ =168.98 K and $T_E$=22.28 K. This indicates a moderately enhanced electronic Sommerfeld coefficient, considerably larger than $\gamma$ = 5.88 mJ mol$_{La}^{-1}$ K$^{-2}$ observed in La$_3$ScBi$_5$. By contrast, The low temperature $C_p/T$-$T^2$ data from 1.8 to 25 K was well described by Debye's description $C/T = \gamma + \beta T^2$, yielding $\gamma^D$ = 43.5 mJ mol$^{-1}$ K$^{-2}$ and $\beta$ = 1.73×10$^{-3}$ J mol$^{-1}$ K$^{-4}$. The two models obtained similar Sommerfeld coefficients, which demonstrated the rationality of data fitting from multiple perspectives, pointing to a slight enhancement in the effective mass and the existence of partial hybrid *f*-electron states[31]. The magnetic specific heat divided by temperature of Pr$_3$ScBi$_5$, illustrated in Fig. 3d as a function of log T, exhibits two distinct features: A jump around $T_N$ and a broad maximum around 10 K. The latter may correspond to the Schottky anomaly arising from the thermal variation of the 4*f* CEF energy levels[32]. The magnetic entropy $S_m$ reaches only 28% × 3Rln2 at $T_N$ and increases up to 3Rln2 by 15 K, which suggests a quasi-doublet ground state. The considerably reduced entropy relative to Rln2 is consistent with the presence of spin fluctuations at temperatures higher than $T_N$, as reflected by the moderately $\gamma^D$ value estimated from paramagnetic regime.

The magnetic susceptibility $\chi(T)$ was measured in an external field of 1 kOe applied along the [2$\bar{1}\bar{1}$0] and [0001] directions, as depicted in Fig. 3e. Both $\chi(T)$ curves exhibit no discernible distinction between zero-field-cooling and field-cooling and feature a sharp peak around 5.1K, indicative of antiferromagnetic (AFM) ordering of the Pr$^{3+}$ moments. Throughout the entire temperature range, the $\chi(T)$ value at $H$ // [2$\bar{1}\bar{1}$0] is one order of magnitude greater than that at $H$ // [0001]. This significant magnetic anisotropy is widespread in the *Ln*$_3$ScBi$_5$ system, indicating that the ordered moments of rare-earth atoms are aligned within the *ab*-plane. The inverse magnetic susceptibility, $\chi^{-1}(T)$, depicted in Fig. 3f, accurately follows the Curie-Weiss law. The estimated effective moments for $H$ // [2$\bar{1}\bar{1}$0] and $H$ // [0001] are 3.786 $\mu_B$ and 3.757 $\mu_B$, respectively, both of which are close to the theoretical value of the free Pr$^{3+}$ ion. The distinct Weiss temperatures of 5.507 K (in-plane) and –106.35 K (out-of-plane) highlight the strong magnetic crystal anisotropy of Pr$^{3+}$, as governed by the CEF[33]. In the *Ln*$_3$ScBi$_5$ system (excluding the non-magnetic element La), where AFM interactions prevail, the Q2D distribution of rare earth atoms within the *ab*-plane strongly influences the AFM order, driven by the synergistic interplay of CEF confinement and magnetic frustration.

**Evolution of Antiferromagnetic order with Field Variation.** Figure 4a displays the $\rho_{\parallel}$ (T) curves, measured below 50 K under various magnetic fields applied along the [2$\bar{1}\bar{1}$0] direction. As magnetic field increases, the resistivity decrease near $T_N$ systematically broadens and shifts to lower temperatures, eventually vanishing at 30 kOe (Fig. S4a, Supporting Information), leading to a pronounced positive MR (discussed further below). $C_p/T$ versus $T$ curves, depicted in Fig.4b, provide additional insights into the thermodynamic aspects of the magnetic correlations and behavior in Pr$_3$ScBi$_5$. The peak associated with $T_N$ shifts to lower temperatures with increasing field and disappears at $H$ = 20 kOe, which is a typical characteristic fingerprint of the AFM transition.

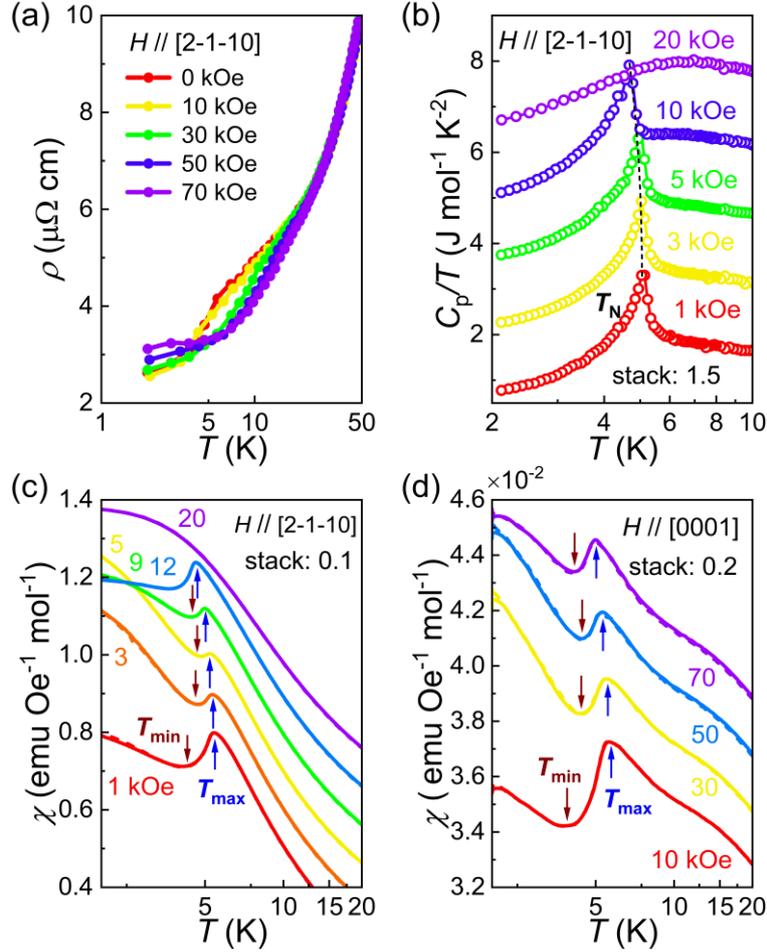

Figure 4. Magnetic order behavior of Pr$_3$ScBi$_5$ measured under varying magnetic fields at low temperatures. (a) Temperature variations of the electrical resistivity for magnetic field parallel to [2$\bar{1}\bar{1}$0] and the current along $c$-axis. (b) Low-temperature $C_p/T$ versus T curves with varied magnetic fields for [2$\bar{1}\bar{1}$0]. (c-d) Temperature-dependent magnetic susceptibility under various magnetic fields for [2$\bar{1}\bar{1}$0] and [0001]. The upward blue arrow and the downward purple arrow denote the local maxima and minima of the magnetic susceptibility curves, respectively.

Consistent with the potential enhancement of spin fluctuations in Pr$_3$ScBi$_5$, the magnetic susceptibility within the basal plane, $\chi(T)$ along the [2$\bar{1}\bar{1}$0] direction, exhibits a broad maximum at $T_{max} > T_N$ and an abrupt drop at $T_N$ upon further cooling, as is corroborated by a sharp peak of d$\chi$/d$T$ (Fig.S4b) and $C_p/T$ versus $T$ data. These intriguing features indicate that, although the static magnetic susceptibility begins to decline at $T_{max}$, the dynamic magnetic susceptibility continues to increase below $T_{max}$ due to spin frustration caused by the Q2D kagome-like structure in $Ln_3$ScBi$_5$, ultimately leading to the emergence of a long-range AFM order[34]. This distinctive magnetic behavior is also observed in other compounds within the $Ln_3$ScBi$_5$ system, indicating a universal influence of spin frustration on magnetic ordering. In addition, at temperatures lower than $T_N$, the $\chi(T)$ curve reaches a local minimum at $T_{min}$ and subsequently exhibits a pronounced upward rise. This increase in $\chi(T)$ curves is attributed to the hyperfine sensitivity of the Pr$^{3+}$ nuclear spins, which arises due to the large hyperfine field generated by the 4f electrons, characterized by a significant Van Vleck magnetic susceptibility[35]. Based on the above analysis, in the $Ln_3ScBi_5$ system, the geometry of the rare earth atoms is causing magnetic frustration even though the rare earth atoms are far enough apart from one another that direct exchange is most likely not present and thus the system relies on RKKY type long-range interactions which typically cause the impact of a geometrically frustrated lattice.

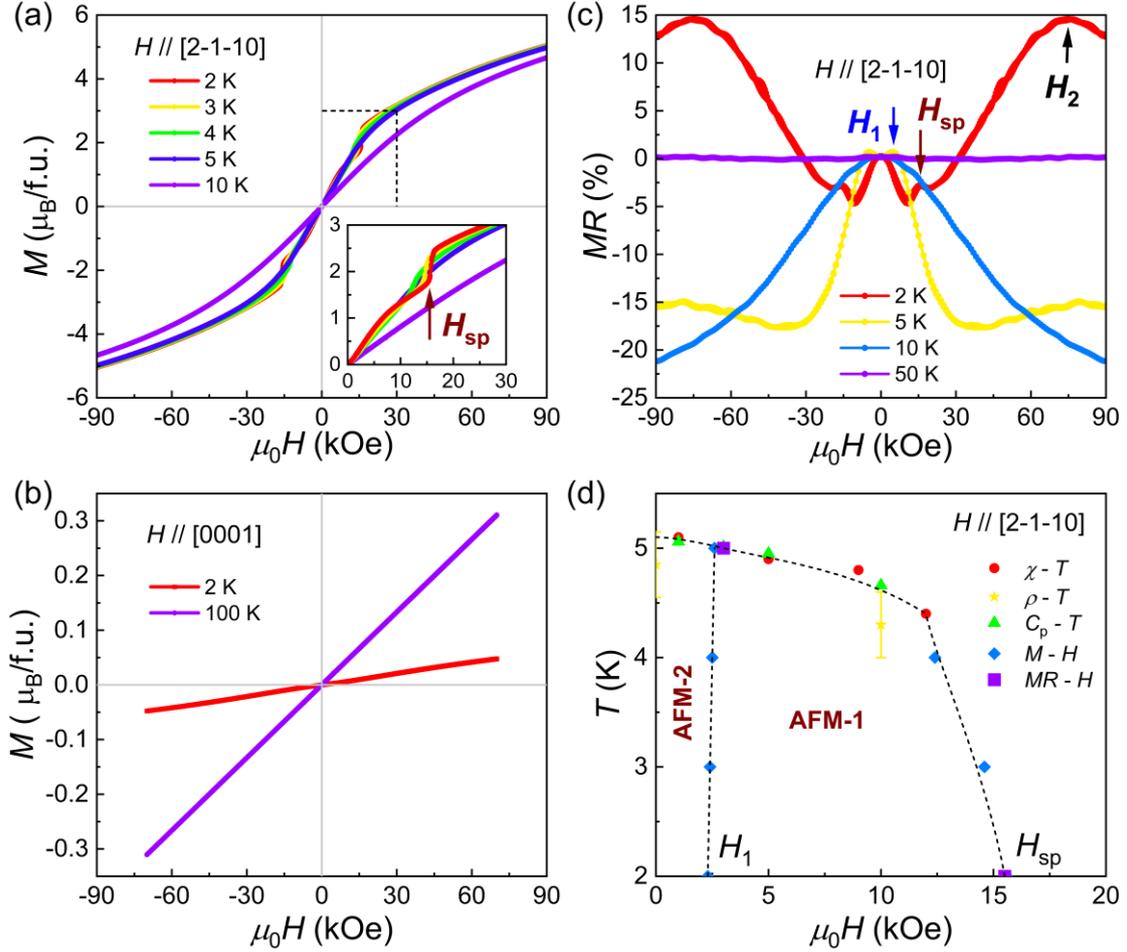

Figure 5. Field-dependent physical properties and phase diagram of $Pr_3ScBi_5$. (a, b) Isothermal magnetizations under various magnetic fields for $H // [2\bar{1}\bar{1}0]$ and $H // [0001]$. (c) Field dependence of magnetoresistance in the magnetic field up to 90 kOe for $H // [2\bar{1}\bar{1}0]$ at various temperatures. (d) Temperature-magnetic field phase diagram of $Pr_3ScBi_5$ for $H // [2\bar{1}\bar{1}0]$.

For $H // [0001]$, the magnetic susceptibility curve is approximately one order of magnitude smaller than that of $H // [2\bar{1}\bar{1}0]$, yet it exhibits a similar peculiar behavior. As the field is applied, the transition at $T_N$ shifts to lower temperatures, persisting even at $H = 70$ kOe (More details in d$\chi$/dT curves can be seen in Fig. S4c, Supporting Information). This observation indicates a strong AFM coupling within the Q1D zigzag chains of the $Ln_3ScBi_5$ system, where spin motion is significantly constrained. Furthermore, we conducted a detailed study of the magnetic properties in the $H // [0\bar{1}10]$ direction, as shown in Fig. S7 of the Supporting Information. The numerical values and trends of the magnetic susceptibility are highly consistent with those in the $H // [2\bar{1}\bar{1}0]$ direction, indicating that $Pr_3ScBi_5$ single crystals exhibit almost no anisotropy within the Q2D kagome plane. This observation holds true for the broader $Ln_3ScBi_5$ system as well. This observation inevitably evokes a classic magnetic structure: the $Pr^{3+}$ moments form a non-collinear AFM arrangement within the Pr triangle of Q2D kagome layers, with spins rotated by 120° relative to one another. Additionally, the Pr moments closest to the [0001] direction are antiparallel, giving rise to three Q1D zig-zag steps of AFM coupling at the Pr sites. In a study utilizing powder neutron diffraction on the isomorphic compound $Nd_3TiSb_5$, a similar 120°AFM order within the kagome plane was reported [36]. This scenario urgently necessitates further analysis and verification through single-crystal neutron diffraction.

**Magnetization, Magnetoresistance and Phase diagram**. The magnetization evolution (Fig.5a-b) under a magnetic field further reveals the uniaxial anisotropic AFM behavior of $Pr_3ScBi_5$. When we apply the field parallel to $[2\bar{1}\bar{1}0]$ direction at 2 K, the in-plane $M(H)$ curve evolves monotonically and exhibits a weak kink near $H_1 \approx$ 3 kOe and a sharp rise near $H_{sp} \approx 15.5$ kOe (More details in d$M$/d$H$ curves can be seen in Fig. S6, Supporting Information), followed by nonlinear increase afterwards. This indicates that the initial AFM-2 state changes to the AFM-1 state at $H_1$, and then to polarized-ferromagnetic (PFM) state via a spin-flop transition at $H_{sp}$. Below $T_N$, $H_1$ and $H_{sp}$ move shift towards lower fields as the temperature rises (Fig.S6, Supporting Information). In contrast, the magnetic moment shows a linear field dependence, as expected for an antiferromagnet under out-of-plane fields (Fig.5b). No indication of magnetization saturation is observed up to 70 kOe for both field orientations. The values of $M(H)$ at 70 kOe at 2 K are 1.38 $\mu_B$/$Pr^{3+}$ ($H // [2\bar{1}\bar{1}0]$) and $1.59 \times 10^{-2}$ $\mu_B$/$Pr^{3+}$ ($H //$ [0001]), which are substantially less than the full moment of free $Pr^{3+}$ ion, suggesting the significant role of the CEF on the ground-state multiplet [28, 37]. These distinct magnetization behaviors offer a further confirm the in-plane arrangement of the Pr magnetic moments, aligning with our previous conjecture about the magnetic structure.

Magnetoresistance (MR) implies some sort of coupling between magnetic order and the electronic structure. In Fig. 5c, we present the calculated MR [MR (%) = 100 × [$\rho_\parallel(H) - \rho_\parallel(H=0)$]/ $\rho_\parallel(H=0)$] at various temperatures, with current directed along the [0001] direction and field aligned parallel to $[2\bar{1}\bar{1}0]$ direction. At $T$ = 2 K, the MR initially exhibits negative MR, and then through the spin flop transformation, it becomes monotonically increasing until $H_2$ = 75 kOe reaches the maximum positive MR of 15%. Conversely, when the temperature rises to 5 K, the spin flop transition vanishes (Fig.S6), and the MR slightly increases below the $H_1$ field before transitioning to negative behavior, ultimately achieving a 15% negative MR at 90kOe. Given the intricate magnetic structure exhibited by $Pr_3ScBi_5$ and the role of spin-dependent electron scattering, further investigations are essential to fully elucidate the underlying mechanisms governing this MR behavior. As illustrated in Fig. 5d, by compiling the critical fields and ordering temperatures derived from multiple measurements, we construct the temperature-magnetic field phase diagram of $Pr_3ScBi_5$ for $H // [2\bar{1}\bar{1}0]$. The phase diagram delineates the AFM ordered phases AFM below 5.1 K and 15.5 kOe, where the AFM1 phase and the AFM2 phase are separated by the field-induced transition marked by the peaks in d$M$/d$H$ and $MR - H$ at $H_1$. The phase boundary lines of the AFM phase and the possible PFM states in other regions are determined jointly by the magnetic transitions from $\chi$ - $T$, $\rho$ - $T$, and $C_p$ - $T$ measurements, as well as the spin flop transitions observed in M - H and MR data. The transitions and anomalies evident in various measurements highlight the intricate interplay between magnetic order arising from the mixing of ground and low-lying singlets and the effects of the applied magnetic field[38, 39]. Investigating the behavior of the quantum critical state when the thermally induced magnetic order in the $Ln_3ScBi_5$ system is driven to zero remains a compelling subject for future research.

## CONCLUSION

The $Ln_3ScBi_5$ family has emerged as one of the most impactful and chemically versatile kagome platforms, characterized by its robust structural framework and tunable physical properties. This research commenced with a thorough evaluation of the currently known $Ln_3ScBi_5$ materials, offering a robust methodology for visualizing structural polymorphs and evaluating their stability boundaries within this family. Simultaneously, we successfully optimized the conditions for flux growth synthesis, yielding a series of high-quality single crystals from the $Ln_3ScBi_5$ family, in which the rare earth atoms exhibited the zig-zig chains typical of Q1D motifs and within the distorted kagome layers of Q2D networks.

We utilized $Pr_3ScBi_5$ as the research medium to investigate the intricate magnetic ground states and distinctive transport phenomena arising from the mixed-dimensional structure of the system. $Pr_3ScBi_5$ exhibits an AFM phase transition at $T_N = 5.1$ K, independent of the magnetic field orientation along the [$2\bar{1}\bar{1}0$] or [0001] crystallographic directions. Our study reveals that the ordered magnetic moments of $Pr^{3+}$ ions are confined within the Q2D kagome planes, displaying minimal in-plane anisotropy. In contrast, a strong AFM coupling is observed within the Q1D zigzag chains, where spin motion is significantly constrained. This distinctive magnetic structural characteristic is commonly observed across the $Ln_3ScBi_5$ family of compounds. Furthermore, the substantial reduction in magnetic entropy at $T_N$ compared to Rln2, along with the observation that AFM phase transitions occur at lower temperatures below the wide peak of magnetic susceptibility, suggests the presence of pronounced spin fluctuations in $Pr_3ScBi_5$. A plausible mechanism within the $Ln_3ScBi_5$ system is that the rare earth atoms are sufficiently spaced to preclude direct exchange interactions and thus the system relies on RKKY type long-range interactions which typically cause the impact of a geometrically frustrated lattice. The $Ln_3ScBi_5$ materials, with its significant structural flexibility, is a promising candidate for the emergence of physical states in mixed-dimensional motifs in bulk materials.

*Acknowledgements.* This work was supported by the National Key R&D Program of China (Grant No. 2024YFA1408400, 2021YFA1400401), the National Natural Science Foundation of China (Grant No. U22A6005, 52271238), the China Postdoctoral Science Foundation under Grant Number 2025M770186, the Center for Materials Genome, and the Synergetic Extreme Condition User Facility (SECUF, https://cstr.cn/31123.02.SECUF). The AI-driven experiments, simulations and model training were performed on the robotic AI-Scientist platform of Chinese Academy of Sciences. And the support was provided by the Research Funds for the Central Universities (N25ZLE007).

Supporting Information:

Stability frontiers and mixed dimensional physics in the Kagome

Intermetallics $Ln_3ScBi_5$ (*Ln*: La−Nd, Sm)

Zhongchen Xu[1, 2, 3], Wenbo Ma[2, 3], Shijun Guo[2, 3], Ziyi Zhang[2, 3], Quansheng Wu[3], Xianmin Zhang[1, #1], Xiuliang Yuan[2, 3, #2], Youguo Shi[2, 3, 4, #3]

[1]*Key Laboratory for Anisotropy and Texture of Materials (Ministry of Education), School of Material Science and Engineering, Northeastern University, Shenyang 110819, China*
[2]*Center of Materials Science and Optoelectronics Engineering, University of Chinese Academy of Sciences, Beijing 100190, China*
[3]*Beijing National Laboratory for Condensed Matter Physics and Institute of Physics, Chinese Academy of Sciences, Beijing 100190, China*
[4]*Songshan Lake Materials Laboratory, Dongguan, Guangdong 523808, China*

To whom correspondence should be addressed.

---

[1] E-mail address: zhangxm@atm.neu.edu.cn
[2] E-mail address: yuanxiuliang8@iphy.ac.cn
[3] E-mail address: ygshi@iphy.ac.cn

Supplemental materials include the following:

1. Table S1 shows crystallographic and structure refinement data of $Ln_3ScBi_5$ ($Ln$ = Pr, Nd, Sm).
2. Table S2 shows atomic coordinates and equivalent isotropic thermal parameters of $Ln_3ScBi_5$ ($Ln$ = Pr, Nd, Sm).
3. Table S3 shows anisotropic atomic displacement parameters ($Å^2$) for $Ln_3ScBi_5$ (Ln = Pr, Nd, Sm).
4. Figure S1 shows characterization of $Pr_3ScBi_5$.
5. Figure S2 shows characterization of $Nd_3ScBi_5$.
6. Figure S3 shows characterization of $Sm_3ScBi_5$.
7. Figure S4 shows the derivative of resistivity ($d\rho/dT$) and magnetic susceptibility ($d\chi/dT$) for $Pr_3ScBi_5$ under various magnetic fields.
8. Figure S5 shows temperature dependence of electrical resistivity for magnetic fields applied parallel to the [0001] direction with current flowing along the c-axis.
9. Figure S6 shows the $dM/dH$ curves for $Pr_3ScBi_5$ were measured at different temperatures for [2$\bar{1}\bar{1}$0].
10. Figure S7 shows Magnetism of $Pr_3ScBi_5$ in an external field parallel to [0$\bar{1}$10].
11. Figure S8 shows the supplementary transport data.

**Table S1.** Crystal data and structure refinement data of $Ln_3ScBi_5$ ($Ln$ = Pr, Nd, Sm) at room temperature, with estimated standard deviations in parentheses.

| Chemical formula | $Pr_3ScBi_5$ | $Nd_3ScBi_5$ | $Sm_3ScBi_5$ |
|---|---|---|---|
| Temperature | | 274(2) K | |
| Formula weight | 1512.60 g/mol | 1522.59 g/mol | 1540.92 g/mol |
| Radiation | | Mo $K\alpha$ 0.71073 Å | |
| Crystal system | | Hexagonal | |
| Space group | | $P6_3/mcm$ | |
| Unit-cell dimensions | $a$ = 9.648(2) Å | $a$ = 9.623(2) Å | $a$ = 9.5332(7) Å |
| | $b$ = 9.648(2) Å | $b$ = 9.623(2) Å | $b$ = 9.5332(7) Å |
| | $c$ = 6.513(2) Å | $c$ = 6.483(2) Å | $c$ = 6.4532(8) Å |
| Volume | 525.0 (3) Å$^3$ | 519.9(3) Å$^3$ | 507.91(10) Å$^3$ |
| Z | | 2 | |
| Density (calculated) | 9.570 g/cm$^3$ | 9.727 g/cm$^3$ | 10.076 g/cm$^3$ |
| Absorption coefficient | 97.635 mm$^{-1}$ | 99.512 mm$^{-1}$ | 103.863 mm$^{-1}$ |
| F(000) | 1226 | 1232 | 1244 |
| $\Theta$ range for data collection | 2.44 to 30.49° | 2.44 to 28.28° | 2.47 to 26.27° |
| Index ranges | -13 <= h <= 13, | -12 <= h <= 12, | -11 <= h <= 9, |
| | -13 <= k <= 13, | -12 <= k <= 12, | -11 <= k <= 11, |
| | -9 <= l <= 9 | -8 <= l <= 8 | -8 <= l <= 8 |
| Independent reflections | 319 [$R_{(int)}$ = 0.0716] | 260 [$R_{(int)}$ = 0.0902] | 209 [R(int) = 0.0738] |
| Structure solution program | | SHELXT 2018/2 (Sheldrick, 2018) | |
| Refinement method | | Full-matrix least-squares on $F^2$ | |
| Refinement program | | SHELXL-2018/3 (Sheldrick, 2018) | |
| Function minimized | | $\Sigma w (F_o^2 - F_c^2)^2$ | |
| Data / restraints / parameters | 319 / 0 / 14 | 260 / 0 / 14 | 209 / 0 / 14 |
| Goodness-of-fit on $F^2$ | 1.081 | 0.778 | 0.930 |
| Final R indices | 311 data; I>2σ(I) | 249 data; I>2σ(I) | 203 data; I>2σ(I) |
| | R1 = 0.0197, | R1 = 0.0195, | R1 = 0.0193, |
| | wR2 = 0.0446 | wR2 = 0.0576 | wR2 = 0.0500 |
| | all data | all data | all data |
| | R1 = 0.0204, | R1 = 0.0201, | R1 = 0.0198, |
| | wR2 = 0.0450 | wR2 = 0.0584 | wR2 = 0.0503 |
| Weighting scheme | w = 1/[σ$^2$(F$_o^2$) + (0.0176P)$^2$+7.1200P] where P=(F$_o^2$+2F$_c^2$)/3 | w = 1/[σ$^2$(F$_o^2$) + (0.0599P)$^2$+1.2064P] where P=(F$_o^2$+2F$_c^2$)/3 | w = 1/[σ$^2$(F$_o^2$) + (0.0389P)$^2$+7.0446P] where P=(F$_o^2$+2F$_c^2$)/3 |

**Table S2.** Atomic coordinates and equivalent isotropic atomic displacement parameters (Å$^2$ ×10$^3$) of $Ln_3ScBi_5$ ($Ln$ = Pr, Nd, Sm) at room temperature.

| Atom | Wyckoff | x/a | y/b | z/c | Occup[a] | $U_{eq}$[b] |
|---|---|---|---|---|---|---|
| **Pr₃ScBi₅** | | | | | | |
| Bi(1) | 6g | 0.73659(4) | 0 | 1/4 | 1.000 | 7.10(15) |
| Bi(2) | 4d | 1/3 | 2/3 | 1/2 | 1.000 | 7.43(16) |
| Pr | 6g | 0 | 0.38225(7) | 1/4 | 1.000 | 8.62(18) |
| Sc | 2b | 0 | 0 | 1/2 | 1.000 | 6.6(7) |
| **Nd₃ScBi₅** | | | | | | |
| Bi(1) | 6g | 0.73583(5) | 0 | 1/4 | 1.000 | 11.5(2) |
| Bi(2) | 4d | 1/3 | 2/3 | 1/2 | 1.000 | 11.7(2) |
| Nd | 6g | 0 | 0.38227(7) | 1/4 | 1.000 | 13.0(2) |
| Sc | 2b | 0 | 0 | 1/2 | 1.000 | 11.0(8) |
| **Sm₃ScBi₅** | | | | | | |
| Bi(1) | 6g | 0 | 0.73441(6) | 1/4 | 1.000 | 8.5(3) |
| Bi(2) | 4d | 2/3 | 1/3 | 1/2 | 1.000 | 8.8(3) |
| Sm | 6g | 0.38265(8) | 0 | 1/4 | 1.000 | 10.1(3) |
| Sc | 2b | 0 | 0 | 1/2 | 1.000 | 10.3(11) |

[a] *Occup*: Occupancy.
[b] $U_{eq}$: one third of the trace of the orthogonalized $U_{ij}$ tensor.

**Table S3.** Anisotropic atomic displacement parameters (Å$^2$) for $Ln_3ScBi_5$ ($Ln$ = Pr, Nd, Sm) at room temperature.

| Lable | U$_{11}$ | U$_{12}$ | U$_{13}$ | U$_{22}$ | U$_{23}$ | U$_{33}$ |
|---|---|---|---|---|---|---|
| **Pr$_3$ScBi$_5$** | | | | | | |
| Pr | 0.0111(3) | 0.00556(17) | 0 | 0.0091(2) | 0 | 0.0063(3) |
| Sc | 0.0091(10) | 0.0046(5) | 0 | 0.0091(10) | 0 | 0.0016(13) |
| Bi(1) | 0.00722(18) | 0.00461(11) | 0 | 0.0092(2) | 0 | 0.0055(2) |
| Bi(2) | 0.00882(19) | 0.00441(10) | 0 | 0.00882(19) | 0 | 0.0047(2) |
| **Nd$_3$ScBi$_5$** | | | | | | |
| Nd | 0.0148(4) | 0.0074(2) | 0 | 0.0125(3) | 0 | 0.0124(4) |
| Sc | 0.0115(12) | 0.0058(6) | 0 | 0.0115(12) | 0 | 0.0099(18) |
| Bi(1) | 0.0109(3) | 0.00637(14) | 0 | 0.0127(3) | 0 | 0.0116(3) |
| Bi(2) | 0.0121(3) | 0.00607(13) | 0 | 0.0121(3) | 0 | 0.0109(4) |
| **Sm$_3$ScBi$_5$** | | | | | | |
| Sm | 0.0073(4) | 0.0049(3) | 0 | 0.0099(5) | 0 | 0.0139(4) |
| Sc | 0.0083(16) | 0.0042(8) | 0 | 0.0083(16) | 0 | 0.014(3) |
| Bi(1) | 0.0076(4) | 0.00381(18) | 0 | 0.0060(3) | 0 | 0.0123(4) |
| Bi(2) | 0.0074(3) | 0.00372(16) | 0 | 0.0074(3) | 0 | 0.0115(5) |

The anisotropic atomic displacement factor exponent takes the form: $-2\pi^2[\ h^2\ a^{*2}\ U_{11} + ... + 2hka^*b^*U_{12}]$

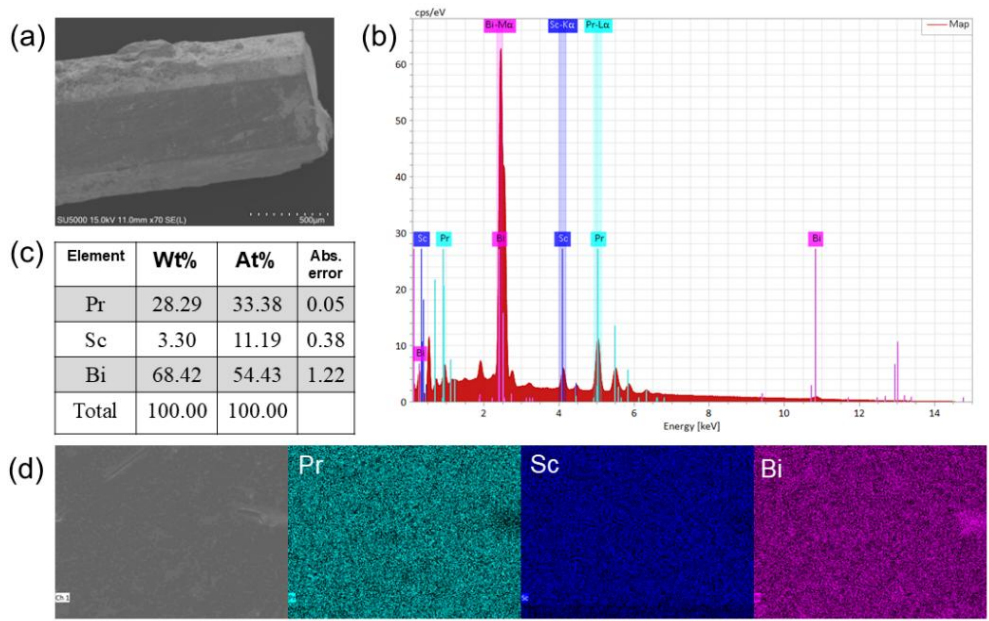

**Figure S1.** Characterization of Pr$_3$ScBi$_5$. (a) SEM image of crystallographic morphology. (c) weight and atomic percentage of Pr, Sc, and Bi atoms. (b), (d) EDX elemental color mapping for Pr, Sc, and Bi for the area, respectively.

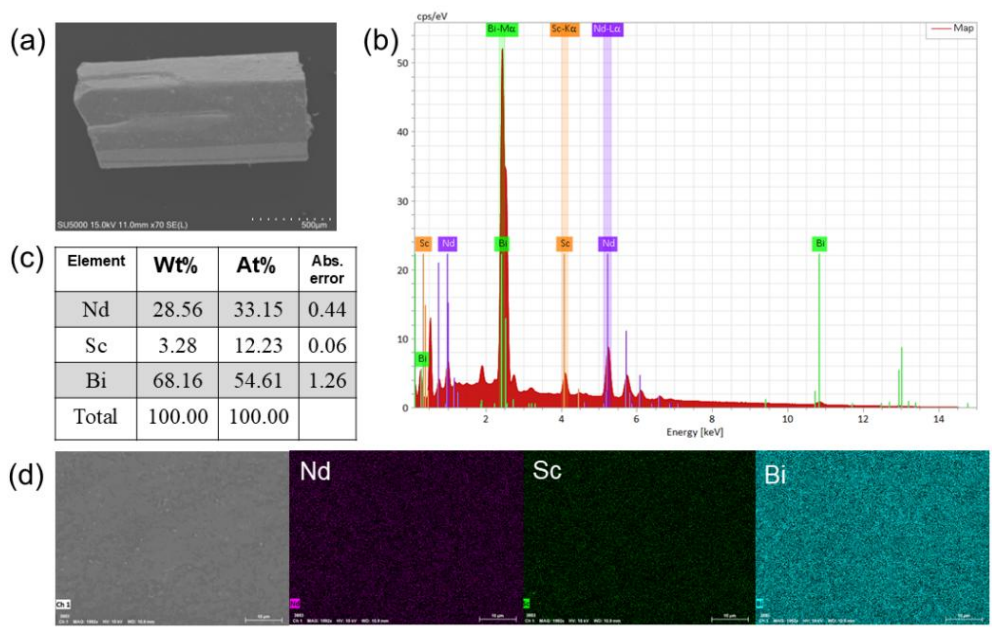

**Figure S2.** Characterization of Nd$_3$ScBi$_5$. (a) SEM image of crystallographic morphology. (c) weight and atomic percentage of Nd, Sc, and Bi atoms. (b), (d) EDX elemental color mapping for Nd, Sc, and Bi for the area, respectively.

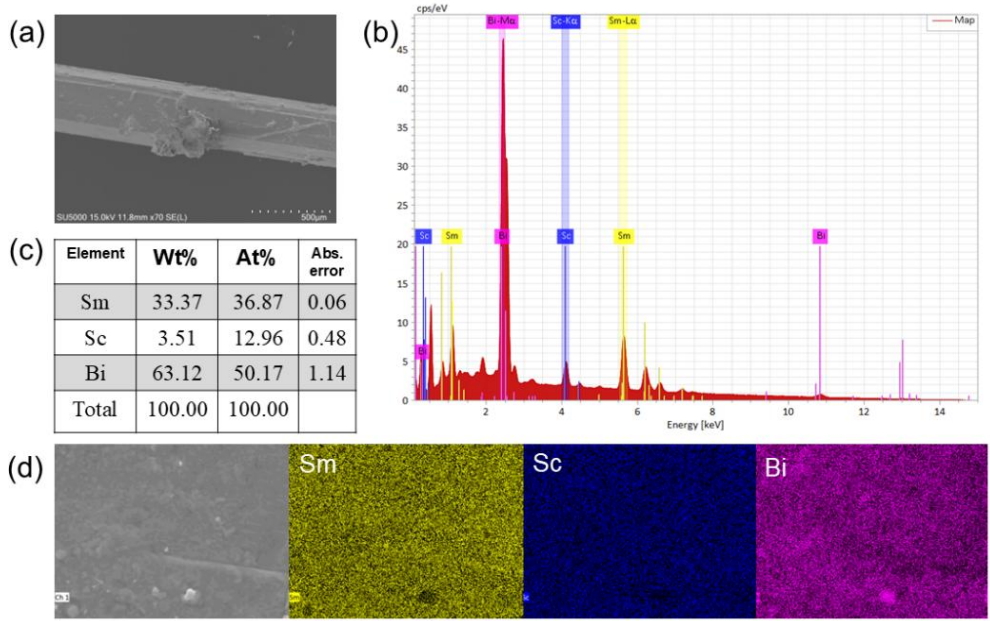

**Figure S3.** Characterization of Sm$_3$ScBi$_5$. (a) SEM image of crystallographic morphology. (c) weight and atomic percentage of Sm, Sc, and Bi atoms. (b), (d) EDX elemental color mapping for Sm, Sc, and Bi for the area, respectively.

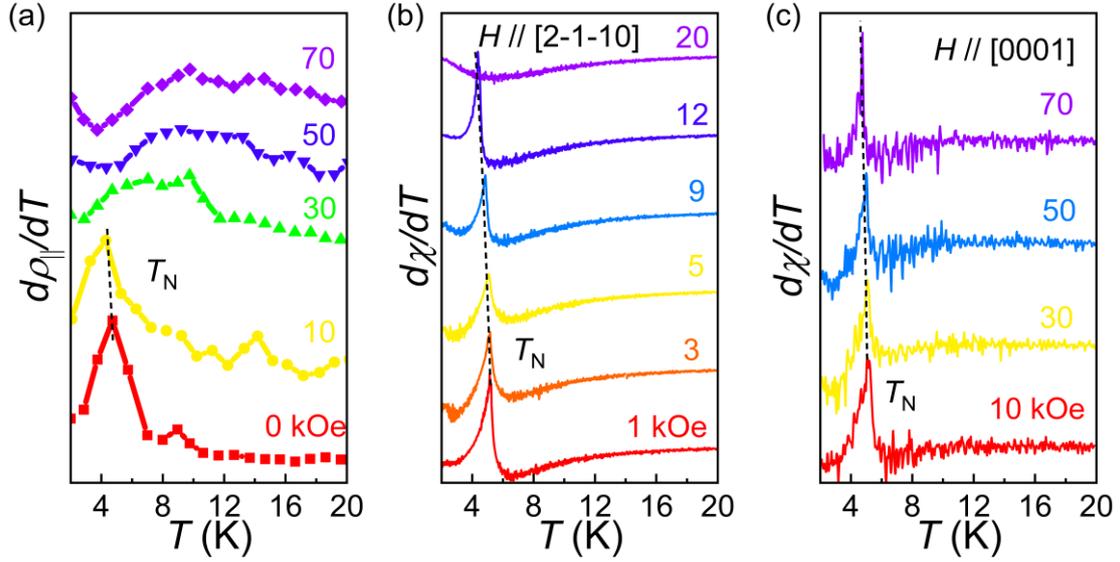

**Figure S4.** The derivative of (a) resistivity ($d\rho/dT$) and (b-c) magnetic susceptibility ($d\chi/dT$) for $Pr_3ScBi_5$ under various magnetic fields. Data points have been vertically shifted for clarity, and short dotted lines are guides to the eye.

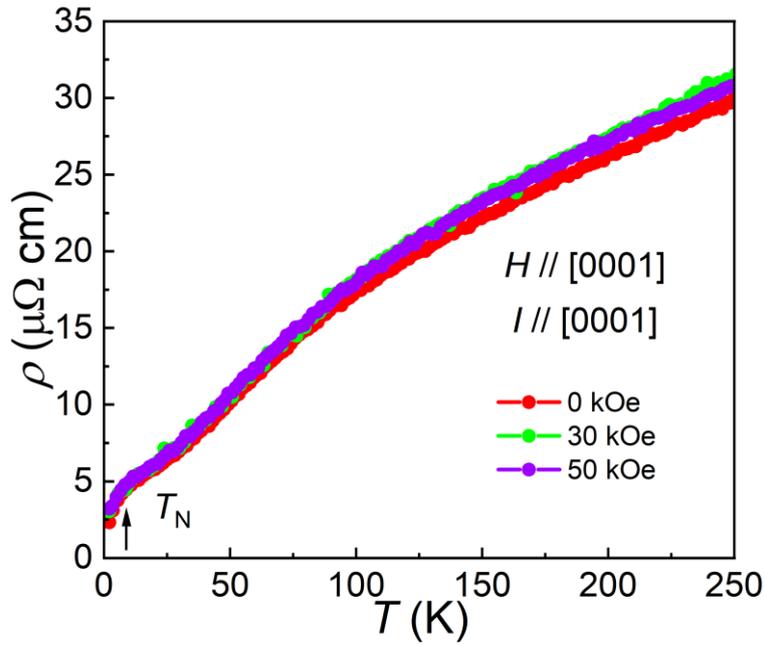

**Figure S5.** Temperature dependence of electrical resistivity for magnetic fields applied parallel to the [0001] direction with current flowing along the c-axis. When the external magnetic field is applied parallel to the [0001] direction, the resistance-temperature curve exhibits a sudden drop around 5 K, indicating the occurrence of a magnetic transition. Additionally, when the field is increased to 50 kOe, this abrupt change persists, consistent with the magnetic susceptibility behavior shown in Figure 4(d).

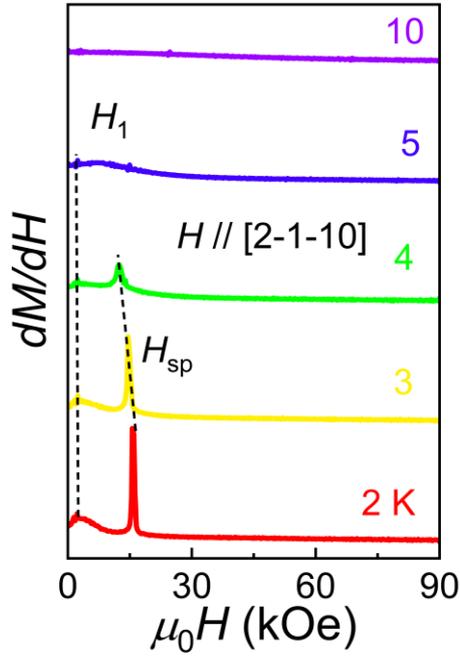

**Figure S6.** The d*M*/d*H* curves for Pr$_3$ScBi$_5$ were measured at different temperatures for [$2\bar{1}\bar{1}0$].

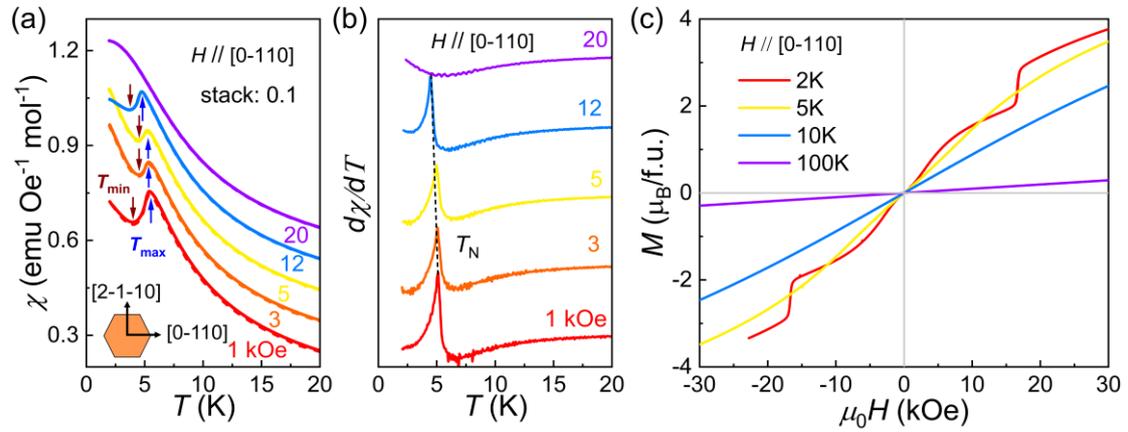

**Figure S7.** Magnetism of Pr$_3$ScBi$_5$ in an external field parallel to [$0\bar{1}10$]. (a) Temperature-dependent magnetic susceptibility under various magnetic fields. (b) The derivative of magnetic susceptibility (d$\chi$/d*T*) under various magnetic fields. (c) Isothermal magnetizations under various magnetic fields.

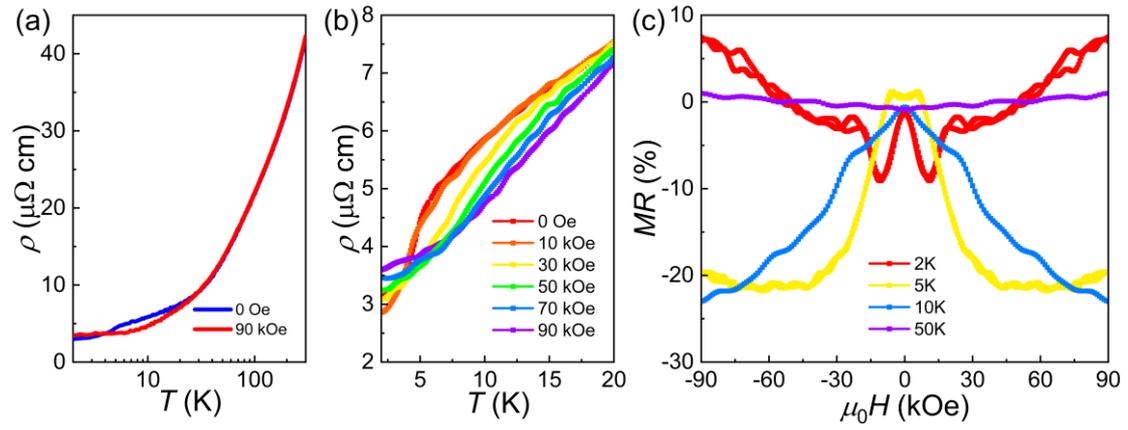

**Figure S8.** Supplementary transport measurements were conducted using different single crystal samples from the same batch as those used for the physical property tests described in the main text.